\documentclass{svmult}

\usepackage{graphicx}
\usepackage{amssymb}
\usepackage{amsmath}
\usepackage{amsfonts}
\usepackage{mathrsfs}
\usepackage{cite}
\usepackage[sort&compress,numbers]{natbib}

\begin{document}
\title{Target search on DNA - effect of coiling}
\author{Michael A. Lomholt and Ralf Metzler}
%\institute{Michael A. Lomholt \at PhyLife - Center for Physical Life Science, University of Southern Denmark, Campusvej 55, 5230 Odense M, Denmark \email{mlomholt@sdu.dk} \and Ralf Metzler \at Institute of Physics \& Astronomy, University of Potsdam, Karl-Liebknecht-Str 24/25, Haus 28, D-14476 Potsdam-Golm, Germany \email{rmetzler@uni-potsdam.de}}
\date{\today}

\maketitle

\abstract{Some proteins can find their targets on DNA faster than by pure
diffusion in the three-dimensional cytoplasm, through the process of facilitated
diffusion: They can loosely bind to DNA and temporarily slide along it,
thus being guided by the DNA molecule itself to the target. This chapter
examines this process in mathematical detail with a focus on including the
effect of DNA coiling on the search process.}

\section{Introduction}

Already in 1968 Adam and Delbr{\"u}ck \cite{adam68} proposed that
diffusion-limited biological processes could be aided by the mechanism of dimensional
reduction. A prime example of this are proteins searching for a specific
location on a long DNA molecule, typically megabases long. In 1970 Riggs
\textit{et al.} \cite{riggs70} observed the very fast association of the
lac repressor protein to its specific operator site on DNA, and Richter
and Eigen suggested in 1974 \cite{richter74} that it was exactly by
exploiting the possibility of one-dimensional diffusion along the DNA that
the lac repressor could be so fast. Detailed mathematical modelling of this
facilitated diffusion process, during which the protein can switch between
three-dimensional diffusion in water and one-dimensional diffusion along the
DNA has been refined in a number of articles by Berg and various coauthors
\cite{berg76,berg77,berg81,berg82,hippel89}. For a more contemporary work on dimensional reduction we refer the reader to \cite{grebenkov22}.

The present chapter examines the effect of DNA coiling on the target search
process of proteins searching for a target on DNA.  The sliding of DNA-binding
proteins was demonstrated in single DNA experiments for both single-strand
DNA binders \cite{sokolov05} and double-strand binders \cite{wang06,hammar12}.  Berg and
Blomberg suggested already in 1977 \cite{berg77} that coiling would speed up
the target localisation by removing some of the correlations that would be
present for a straight DNA molecule.  The speed-up by DNA coiling has been
confirmed experimentally by Gowers and Halford exploiting supercoiling of
DNA in \cite{gowers03}, and later by optical tweezers experiments by Broek
{\it et al.} \cite{broek08}. The theory of the effect of DNA coiling that is
presented in this chapter is focused on modelling the latter experiment. It
builds on the facilitated diffusion model by Berg and Ehrenberg \cite{berg82}
for straight DNA and it constitutes a significantly more detailed version
of the mathematical theory presented in \cite{lomholt09}.

The account begins with a re-derivation of Smoluchowski's 1916 result for
diffusion-limited reactions with spherical targets \cite{smoluchowski16}
in Section \ref{sec:smoluchowski}. This is the target localisation rate that
dimensional reduction helps in surpassing. Section \ref{sec:smoluchowski}
also introduces, for this simple case, some of the mathematical apparatus
that the remaining sections use. Next the facilitated diffusion process
is presented from the point of view of diffusion along the DNA in Section
\ref{sec:facilitated}. The specific theory for straight and coiled DNA
then follows in Sections \ref{sec:straight} and \ref{sec:coiled}. In Section
\ref{sec:intra} the mechanism of intersegmental transfers is briefly discussed
before ending with conclusions in Section \ref{sec:conclusion}. Additionally,
the Matlab scripts used for calculating target localisation rates and creating
the figures in this chapter can be found at \cite{coiledDNA}.

\section{Smoluchowski result}
\label{sec:smoluchowski}

We will first revisit the case of diffusing proteins reacting with a spherical target at the origin as first studied by Smoluchowski in 1916 \cite{smoluchowski16}. To find out how fast a protein can find the target we will study the diffusion-limited binding to the target. Our dynamical quantity is the three-dimensional volume density $n=n(r,t)$. We assume spherical symmetry so the density only depends on the radial distance $r$ to the origin and time $t$. The dynamics follows the diffusion equation
\begin{equation}
    \frac{\partial}{\partial t}n(r,t)=D_{\rm 3d} \frac{1}{r^2}\frac{\partial}{\partial r}\left(r^2\frac{\partial}{\partial r}n(r,t)\right),
\label{eq:diff3d}
\end{equation}
where $D_{\rm 3d}$ is the protein diffusion constant. The proteins react with the target placed at the origin whenever they are within a distance $b$ of it (measured from the centre of the protein to the centre of the target). This means that we have an absorbing boundary condition at $r=b$,
\begin{equation}
    n(r=b,t)=0.
\end{equation}
Far away from the target we will assume that the density approaches a constant value $n_{\rm bulk}$, meaning that we have the second boundary condition
\begin{equation}
    \lim_{r\to\infty}n(r,t)=n_{\rm bulk}.
\end{equation}

To solve the diffusion equation we Laplace transform the density $n(r,t)$,
\begin{equation}
    n(r,u)=\int_0^\infty dt\; e^{-u t} n(r,t).
\end{equation}
Note that we indicate a Laplace transformed quantity by a change of the variable to Laplace time $u$. When Laplace transforming the diffusion equation it is converted into the ordinary differential equation
\begin{equation}
u n(r,u)-n_{\rm bulk}=D_{\rm 3d} \frac{1}{r^2}\frac{\partial}{\partial r}\left(r^2\frac{\partial}{\partial r}n(r,u)\right),
\end{equation}
where we assumed that initially the concentration is constant $n(r,t=0)=n_{\rm bulk}$. The general solution to this equation is
\begin{equation}
    n(r,u)=\frac{n_{\rm bulk}}{u}+\frac{A(u)}{r}e^{\sqrt{u/D_{\rm 3d}}r}+\frac{B(u)}{r}e^{-\sqrt{u/D_{\rm 3d}}r},
\end{equation}
where $A$ and $B$ are integration constants that can depend on $u$. These can be identified via the two boundary conditions, leading to the unique selection
\begin{equation}
    n(r,u)=\frac{n_{\rm bulk}}{u}\left(1-\frac{b}{r}e^{-\sqrt{u/D_{\rm 3d}}(r-b)}\right).
\end{equation}

The current density along the radial direction is $-D_{\rm 3d}\partial n/\partial r$. The current $j$ of proteins into the target is minus this current times the surface area. In Laplace space we find
\begin{align}
j(u)&=4\pi b^2 D_{\rm 3d}\left.\frac{\partial n(r,u)}{\partial r}\right|_{r=b}\nonumber\\
&=4\pi D_{\rm 3d} b n_{\rm bulk}\left(\frac{1}{u}+\frac{b}{\sqrt{D_{\rm 3d}u}}\right).
\label{eq8}
\end{align}
Transforming back to real time this gives
\begin{equation}
j(t)=4\pi D_{\rm 3d} b n_{\rm bulk}\left(1+\frac{b}{\sqrt{\pi D_{\rm 3d}t}}\right).\label{eq:jt}
\end{equation}
From this solution for the current into the target we see that it reaches
the stationary value $j_{\rm stat}=4\pi D_{\rm 3d} b n_{\rm bulk}$ at times
$t\gg b^2/D_{\rm 3d}$. With inverse mobility $1/D_{\rm 3d}$ and with the
square of the target radius $b$, this characteristic time increases, as
expected.

Let us for the moment assume that we are in the eventual stationary situation
with a constant flux of proteins into the target. Then at any given moment
there is a probability density $j_{\rm stat}$ of a protein arriving at
the target. If we look at the waiting time $\tau$ before the next protein
arrives, then this will be exponentially distributed with rate parameter
$j_{\rm stat}$, i.e., the average waiting time reads
\begin{equation}
\langle \tau\rangle=j_{\rm stat}^{-1}.
\end{equation}
This equation converts between the macroscopic view of the equations for
the density of proteins and the microscopic view of the arrival of the
first protein to the target. When $\langle \tau\rangle\gg b^2/D_{\rm 3d}$,
i.e., when $n_{\rm bulk}\ll b^{-3}$, then the transient contribution (the
second term) in Eq. (\ref{eq:jt}) is unimportant for the statistics of the
protein binding. In the following we will assume that we are at such dilute
concentrations of proteins such that we only need to study the stationary
regime to know the binding statistics. Furthermore, we will remove the
dependence on concentration by defining the rate constant for the binding
reaction as
\begin{equation}
k_{\rm on}\equiv \frac{j_{\rm stat}}{n_{\rm bulk}}=4\pi D_{\rm 3d} b.
\label{eq:smolres}
\end{equation}
This is indeed Smoluchowski's result \cite{smoluchowski16}.

The experimental motivation for developing another model than
the above for DNA binding proteins comes from in vitro experiments on the
Lac repressor \cite{riggs70} where a value $k_{\mathrm{on}}\approx
10^{10} /[(\mathrm{mol}/\mathrm{l})\cdot\mathrm{sec}]$ was measured. We
can compare this with Smoluchowski's theory for spherical targets by
estimating its prediction for $k_{\mathrm{on}}$. For a typical diameter
of a transcription factor of $2R=5\,{\rm nm}$ by Stokes formula we obtain
$D_{\mathrm{3d}}=6\pi\eta R \approx 10^2 \mu\mbox{m}^2/\mbox{s}$ where we
used the viscosity of water $\eta\approx 10^{-3}\,{\rm Pa}\cdot{\rm s}$. If
we say the protein has to situate itself on the DNA within a precision
matching the length of one base-pair it corresponds to a target of size $2
b\approx 0.3\,$nm. Consequently we find $k_{\mathrm{on}}\approx 0.2\,\mu{\rm
m}^3/{\rm s}\approx 10^8/[(\mathrm{mol}/\mathrm{l})\cdot\mathrm{sec}]$. That
some proteins can operate faster than this upper diffusion limit for spherical
targets motivates the facilitated diffusion model in the next section. We note that the considerations here apply to in vitro situations. In vivo conditions will be different. For instance, a much reduced value for $D_{\rm 3d}$, compared with the estimate in water above, was found inside living \textit{Escherichia coli} in \cite{elf07}.

\section{Facilitated Diffusion along DNA}
\label{sec:facilitated}

\begin{figure}[t]
\begin{center} 
\includegraphics[width=11.8cm]{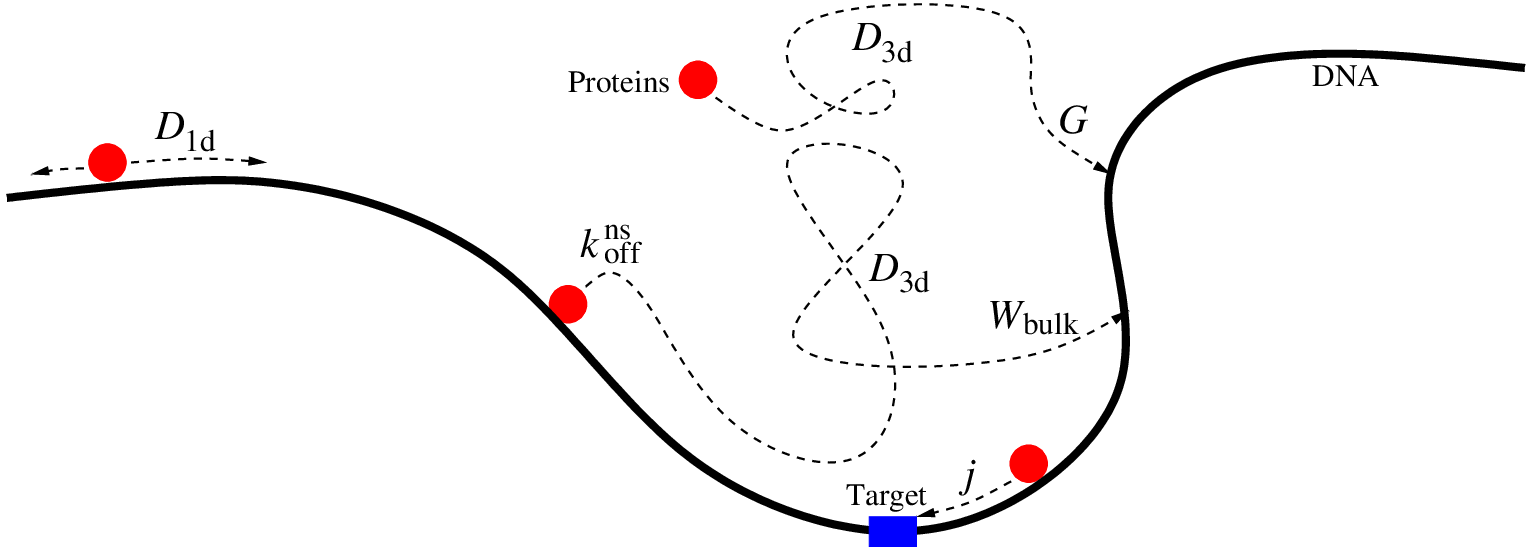}
\end{center}
\caption{Schematic illustration of the different terms in Eq. (\ref{eq:themodel}).
}
\label{fig:dynamics}
\end{figure}

The faster search can be explained by the fact that proteins can bind
loosely ("non-specifically" \cite{ptashne02,stormo98,buchler03}) to the DNA, and then
slide along the DNA diffusively, searching for the target. To model this
process mathematically we will formulate it as a diffusion process along
the DNA. Thus our central dynamic quantity $n(x,t)$ is now the density of
loosely bound proteins per length of the DNA at position $x$, where $x$
is the distance along the DNA contour with the origin defined to be at
the target. We label the diffusion constant along the DNA $D_{\rm 1d}$
to distinguish it from the diffusion constant $D_{\rm 3d}$ for unbound
proteins. The diffusion equation governing $n(x,t)$ then has several extra
terms as compared with Eq. (\ref{eq:diff3d}) (see Fig. \ref{fig:dynamics} for an illustration of this model),
\begin{eqnarray}
&&\hspace*{-0.4cm}\frac{\partial n(x,t)}{\partial t}=\left(D_{\rm 1d}\frac{\partial^2}{ \partial x^2}-k^{\rm ns}_{\rm off}\right)n(x,t)-j(t)\delta(x)+G(x,t)\nonumber\\
&&+k^{\rm ns}_{\rm off}\int_{-\infty}^\infty d x'\int_0^t d t'\,W_{\rm bulk}(x-x',t-t')n(x',t').
\label{eq:themodel}
\end{eqnarray}
Here $k^{\rm ns}_{\mathrm{off}}$ is the dissociation rate from the DNA of
the non-specifically bound proteins; $j(t)$ is, as in the previous section,
the number of proteins per time binding to the target site located at $x=0$
and they are thus removed; $W_{\rm bulk}(x-x',t-t')$ is the joint probability density
that a protein rebinds on the DNA at the point $x$ and at the time $t$ given that it has fallen of the DNA
at $x'$ at time $t'$. The remaining term, $G(x,t)$, accounts for the proteins
that bind to the DNA at time $t$ and position $x$ without previously having
been bound between time zero and time $t$. We will argue later that details of
$G(x,t)$ (such as if it is uniform in $x$) will not matter for dilute protein
concentrations. We are assuming that any protein diffusing across the target
will be absorbed by it, imposing the condition $n(x=0,t)=0$. This
condition determines the current $j(t)$ of proteins binding to the target. We note that in this model the proteins can only find the target by first binding loosely to the DNA, and then diffusing one-dimensionally along the DNA into the target.

In the above we made a number of assumptions to simplify the interplay
between one-dimensional diffusion along the DNA and diffusion in the bulk
around the DNA: We assumed that the DNA is long enough that we can take the
end points to infinity when considering the diffusion to the target, i.e.,
we can integrate to infinity in the convolution integral containing $W_{\rm
bulk}$. This assumption should hold, if the absorption at the target only
depletes proteins at a distance from the target that is small compared with the
distance to the end points. In the next section we will introduce an effective
sliding length that provides a length scale for this depletion. Another
assumption is that there is no dependence between excursion events to the
bulk. A static coiled configuration of the DNA could lead to such dependence,
since two excursions following each other could bring the protein back and
forth between two parts of the DNA that are far away from each measured along
the DNA with high probability. The assumption should hold for straight or
quickly fluctuating DNA though. Assuming we are in either of these situations
the kernel $W_{\rm bulk}(x-x',t-t')$ is assumed to be homogeneous in space
and time too, i.e., the distance and duration of excursions in the bulk do
not depend on the initial location or time. Additionally, we assume that the
diffusion along the DNA is homogeneous, i.e., it does not depend on the local
sequence of nucleotides. See for instance \cite{slutsky04,benichou09} for
discussions of how sequence heterogeneity could be a barrier for fast search,
but also for a mechanism with which proteins can circumvent this barrier. Finally, we assume that the targets extension along the DNA length is short compared with typical sliding lengths along the DNA. This implies that the possibility that the protein can bind to the DNA directly on top of the target via three-dimensional diffusion in the bulk does not matter for the search time, and we therefore consider only the possibility that the protein can find the target by sliding along the DNA.

As in the previous section, we want to find the stationary flux into
the target $j_{\rm stat}$, and we will use the same technique of Laplace
transforming, replacing time $t$ with the Laplace variable $u$. Furthermore,
we will Fourier transform along the DNA:
\begin{equation}
n(q,u)=\int_{-\infty}^\infty dx\; e^{-i q x} n(x,u),
\end{equation}
similarly indicating the transform by replacing the position coordinate $x$ with wavenumber $q$. Applying the two transforms Eq. (\ref{eq:themodel}) becomes
\begin{equation}
\frac{1}{W(q,u)}n(q,u)-n(q,t=0)=G(q,u)-j(u),
\label{eq:Laplace}
\end{equation}
where
\begin{equation}
\frac{1}{W(q,u)}=u+D_{\rm 1d} q^2+k^{\rm ns}_{\rm off}[1-W_{\rm bulk}(q,u)].
\end{equation}
To extract the long time behaviour we rewrite Eq. (\ref{eq:Laplace}) to arrive at
\begin{equation}
n(q,u)=n^{\rm ns}(q,u)-j(u)W(q,u),
\label{eq:densities}
\end{equation}
where we have introduced
\begin{equation}
n^{\rm ns}(q,u)=W(q,u)[G(q,u)+n(q,t=0)].
\end{equation}
$n^{\rm ns}(x,t)$ is the density of proteins on the DNA in the absence of the
sink at the target. For this situation we would expect equilibration to occur,
such that eventually we will have a constant density $n^{\rm ns}_{\rm eq}$, since without the sink there is no inhomogeneity along the DNA. In
the following we will neglect all dependence on the initial distribution of
proteins by assuming we are in this equilibrium limit of loose binding to
the DNA. Due to the linearity of our equations we expect this assumption to
hold in the dilute limit, since the equilibration time should be independent
of the protein concentration, while the search time will increase as
concentration is lowered \cite{sokolov05}. In the Laplace domain this means that $n^{\rm
ns}(x,u)\sim n^{\rm ns}_{\rm eq}/u$, where $\sim$ indicates that we are looking
at the behaviour at long times, which by Tauberian theorems \cite{hughes95} is
equivalent to studying the limit of $u$ going to zero. Equations (\ref{eq8})
and (\ref{eq:jt}) provide an example of this, where it can be found that
the term that dominates at small $u$ converts to a term that dominates at
large $t$, and vice versa for the other term.

We can find the current $j(u)$ by inverse Fourier transforming
Eq. (\ref{eq:densities}) at $x=0$ using the condition of a sink at the target,
$n(x=0,t)=0$, to arrive at
\begin{equation}
j(u)=\frac{n^{\rm ns}(x=0,u)}{W(x=0,u)}. 
\end{equation}
To proceed we will assume that the value $W(x=0,u=0)$ is finite and non-zero,
which will be the case for the three-dimensional diffusion problems we study
here with the bulk space stretching far in all directions, leading to long
tailed distributions for the bulk excursions. With this assumption we find
the limiting behaviour at small $u$: $j(u)\sim k_{\rm 1d}n^{\rm ns}_{\rm
eq}/[u W(x=0,u=0)]$, i.e., that the current becomes stationary at long times
with value
\begin{equation}
j_{\rm stat}\sim k_{\rm 1d}n^{\rm ns}_{\rm eq},
\end{equation}
where $k_{\rm 1d}=1/W(x=0,u=0)$, or explicitly
\begin{equation}
\label{int}
k_{\rm 1d}^{-1}=\int_{-\infty}^\infty \frac{d q}{2\pi}
\frac{1}{D_{\rm 1d}q^2+k_{\rm off}^{\rm ns}(1-\lambda_{\rm bulk}(q))},
\end{equation}
with the distribution $\lambda_{\rm bulk}(x)=W_{\rm bulk}(x,u=0)$ of the
relocation lengths along the DNA contour of the three-dimensional bulk
excursions.

To proceed and calculate the integral in Eq. (\ref{int}) we need to know
$\lambda_{\rm bulk}(q)$. There is, however, a situation we can discuss already
without a detailed form of $\lambda_{\rm bulk}(q)$: If the density $\lambda_{\rm bulk}(x)$
has its probability mass spread out sufficiently such that a protein is unlikely to rebind to a point where
it was previously bound to the DNA, then for the purpose of calculating
the integral in Eq. (\ref{int}) we can effectively consider $\lambda_{\rm
bulk}(q)$ to vanish for $q\ne 0$ (it will always be unity for $q=0$ by
normalisation). We remark that the integrand will be singular at $q=0$. But the overall three dimensional geometry of the bulk means that the rate of target finding will be finite as in Section \ref{sec:smoluchowski}. Thus the singularity will be soft enough for the integral to converge (see the discussion above Eq. (\ref{eff_slide}) for the explicit case of straight DNA). Assuming we are in this situation, where $\lambda_{\rm bulk}(q)$ does not contribute significantly
to the integral, we discard $\lambda_{\rm bulk}(q)$ and then integrate analytically to find
\begin{equation}
k_{\rm 1d}=2\sqrt{D_{\rm 1d}k^{\rm ns}_{\rm off}}.
\label{eq:reactlimit1}
\end{equation}
If we multiply this result on both sides with $n^{\rm ns}_{\rm eq}$ it can be
rewritten as
\begin{equation}
j_{\rm stat}=2 l_{\rm sl} k^{\rm ns}_{\rm off} n^{\rm ns}_{\rm eq}\label{eq:reactlimit2}.
\end{equation}
To understand this expression, first note that $k^{\rm ns}_{\rm off} n^{\rm
ns}_{\rm eq}$ is the rate of unbinding events per length of the DNA. In the
equilibrated situation this is equivalent to the rate of rebinding events per
length. If such a rebinding happens within a distance of $l_{\rm sl}$ on either
side of the target, then the protein will be likely to find the target, thus
resulting in Eq. (\ref{eq:reactlimit2}) for the rate of target localisation.

In the previous section we preferred to express results in terms of $k_{\rm
on}=j_{\rm stat}/n_{\rm bulk}$, where $n_{\rm bulk}$ was the density per
volume in the bulk, i.e., the density of non-bound proteins in the water
around the DNA. At equilibrium of non-specific binding the two concentrations
are related by the non-specific binding constant per length of DNA, which
is $K_{\rm ns}=n^{\rm ns}_{\rm eq}/n_{\rm bulk}$. Using this we can write
\begin{equation}
k_{\rm on}=k_{\rm 1d}K_{\rm ns}.
\label{eq:konk1d}
\end{equation}
Note that if a significant fraction of the proteins are bound to the DNA it
becomes necessary to distinguish the bulk concentration of proteins $n_{\rm
bulk}$ from the total amount of proteins per volume $n_{\rm total}$. If $l_{\rm
DNA}$ is the length of the DNA divided by the volume of the surrounding water,
then the concentrations are related by
\begin{equation}
{n}_{\rm total}={n}_{\rm bulk}+l^{\rm total}
_{\rm DNA}n^{\rm ns}_{\rm eq}=(1+l^{\rm total}
_{\rm DNA}K_{\rm ns}){n}_{\rm bulk}.
\end{equation}
In terms of $n_{\rm total}$ the target localisation rate can therefore be written as
\begin{equation}
j_{\rm stat}=\frac{k_{\rm on}}{1+K_{\rm ns}l_{\rm DNA}^{\rm total}}n_{\rm total}.
\end{equation}
In the following two sections we will aim at calculating $k_{\rm on}$ first for straight and then coiled DNA.

\section{Straight DNA}
\label{sec:straight}

\begin{figure}[t]
\begin{center}
\includegraphics[width=7.8cm]{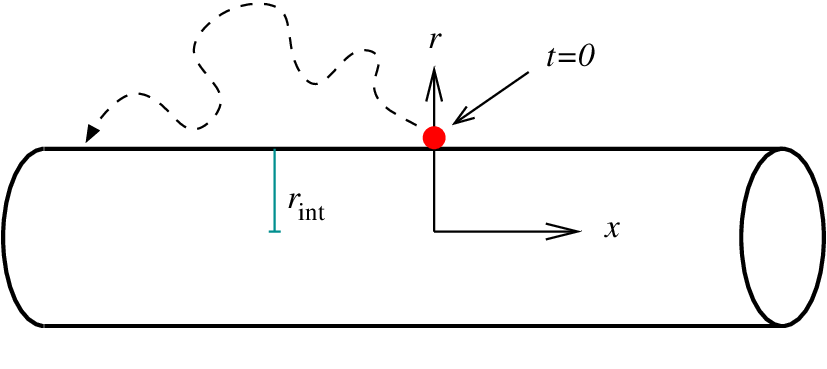}
\end{center}
\caption{Schematic of the bulk excursion of a protein being released at the cylindrical DNA-interaction surface. This process is modelled by the diffusion equation (\ref{eq:bulkdiff}) and the boundary condition (\ref{eq:bulkbound}).}
\label{fig:straight}
\end{figure}

In this section we will obtain $W_{\rm bulk}$ for the case when the DNA is completely straight. This can be achieved, e.g., by stretching the DNA in an optical tweezers setup \cite{pant04,broek08} or in microfluidic setups \cite{krog18b}. We can then consider the problem as a diffusion problem around a cylinder, which is a problem that is well known in the literature, see for instance \cite{carslaw59,chechkin09,chechkin11}. Here we provide a derivation tailored towards obtaining $W_{\rm bulk}(x,t)$. We first determine the probability density $P$ for a single protein to be at a position in space at a certain time $t$. In this cylindrical geometry, the DNA is situated along the $x$-axis, and the protein will interact with it when it is at a distance $r=r_{\rm int}$ from the $x$-axis (see Fig. \ref{fig:straight}). For the directions orthogonal to the $x$-axis we use polar coordinates with $r$ the radial coordinate. Since we are only interested in the value of $x$ at which the protein rebinds to the DNA, we will drop the dependence on the polar angle and $P$ is then rotationally symmetric around the $x$-axis. Thus we have $P=P(x,r,t)$, and the three-dimensional diffusion equation can be written as
\begin{equation}
\frac{\partial {P}}{\partial t}=D_{\rm 3d}\left(\frac{\partial^2 }{\partial x^2}+\frac{1}{r}\frac{\partial}{\partial r}r\frac{\partial}{\partial r}\right){P},
\label{eq:bulkdiff}
\end{equation}
with the diffusion constant $D_{\rm 3d}$. The boundary condition we use at $r=r_{\rm int}$ can be expressed as the probability flux density in the radial direction,
\begin{equation}
-\left. D_{\rm 3d}\frac{\partial {P}}{\partial r}\right|_{r=r_{\rm int}}=-\left.\frac{k^{\rm ns}_{\rm on}}{2 \pi r_{\rm int}}{P}\right|_{r=r_{\rm int}}+\frac{\delta(x)}{2 \pi r_{\rm int}}\delta(t).
\label{eq:bulkbound}
\end{equation}
On the right-hand side the first term represents the probability rate of the protein loosely binding to the DNA per surface area of the cylinder. $k^{\rm ns}_{\rm on}$ is a rate constant with units of area per time. The second term represents the protein release from the boundary at $x=0$ at time $t=0$. Note that consistency with non-specific equilibration, $n^{\rm ns}_{\rm eq}k^{\rm ns}_{\rm off}={n}_{\rm bulk}k^{\rm ns}_{\rm on}$, means that $k^{\rm ns}_{\rm on}=K_{\rm ns} k^{\rm ns}_{\rm off}$. That the protein enters the bulk via the boundary condition means that we will take the bulk to be empty initially: $P(x,r,t=0)=0$. Additionally we assume that $P$ vanishes infinitely far away from the spatial origin at all times.

To solve Eq. (\ref{eq:bulkdiff}) we Laplace transform the time $t$ as usual and also Fourier transform along the $x$-axis again to convert the derivatives with respect to $x$ to factors of $iq$, i.e.,
\begin{equation}
u P(q,r,u)=D_{\rm 3d}\left(-q^2+\frac{1}{r}\frac{\partial}{\partial r}r\frac{\partial}{\partial r}\right){P}(q,r,u).
\label{eq:bulkdiffq}
\end{equation}
The general solution of this equation is given in terms of modified Bessel functions $I_n$ and $K_n$ as
\begin{equation}
    P(q,r,u)=A I_0({\bar q} r)+B K_0({\bar q} r),
\end{equation}
where ${\bar q}=\sqrt{q^2+u/D_{\rm 3d}}$. The boundary condition that $P$ vanishes infinitely far from the origin  tells us that $A=0$. Inserting the term with $B$ into the remaining boundary condition (\ref{eq:bulkbound}) gives
\begin{equation}
   B=\frac{1}{k^{\rm ns}_{\rm on} K_0({\bar q}r_{\rm int}) + 2 \pi r_{\rm int} D_{\rm 3d}{\bar q} K_1({\bar q} r_{\rm int})}.
\end{equation}

Now we have arrived at a result that we can use to calculate $W_{\rm bulk}$, since it is simply given by $W_{\rm bulk}(x,t)=k^{\rm ns}_{\rm on} P(x,r=r_{\rm int},t)$, i.e., the probability that the protein binds at the boundary at position $x$ and time $t$. This produces the result
\begin{equation}
W_{\mathrm{bulk}}^{\rm cyl}(q,u)= \left(1+\frac{2\pi D_{\rm 3d}{\bar q}r_{\rm int}K_1({\bar q} r_{\rm int})}{k^{\rm ns}_{\rm on}K_0({\bar q} r_{\rm int})}\right)^{-1}.
\label{eq:Wbulkcyl}
\end{equation}
The superscript cyl indicates that this is the solution when the DNA is forming a straight cylinder.

\begin{figure}[t]
%\sidecaption
\begin{center}
\includegraphics[width=8cm]{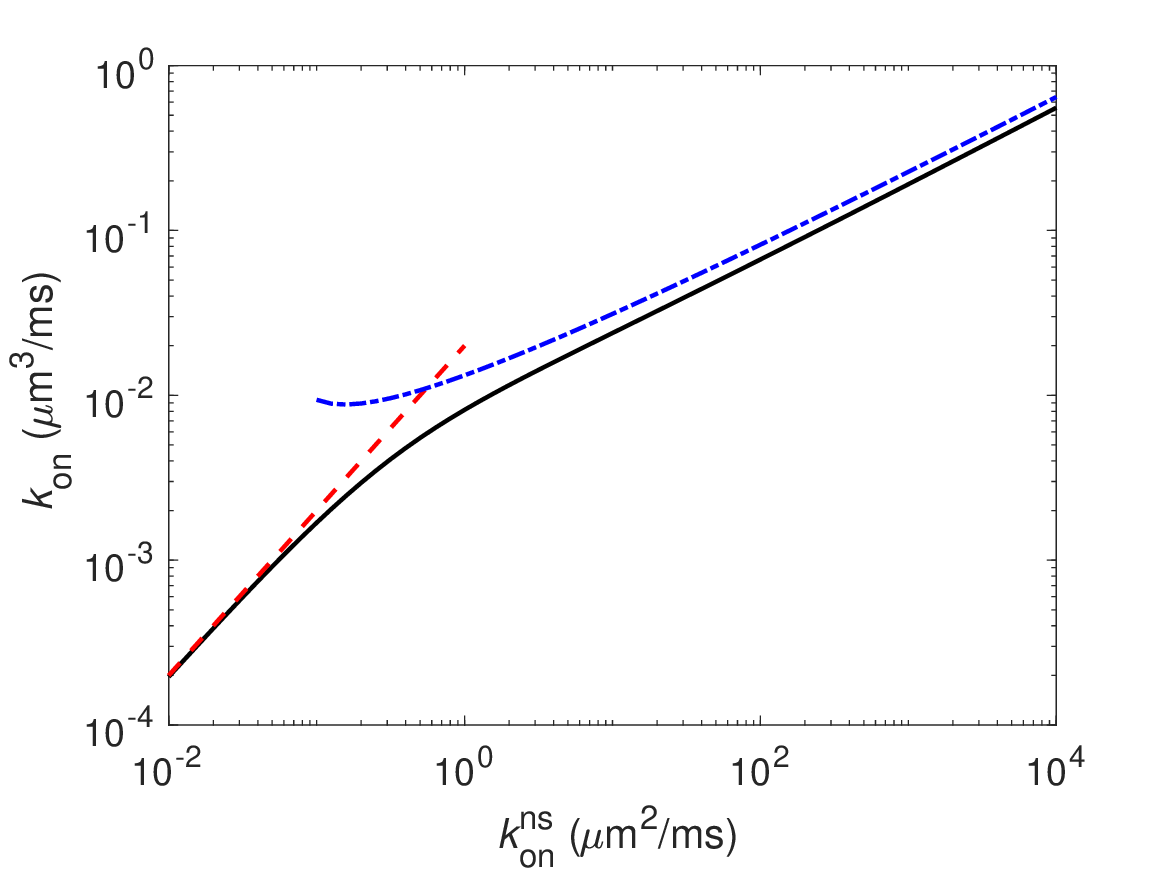}
\end{center}
\caption{Plot of $k_{\rm on}$ as a function of $k_{\rm on}^{\rm ns}$ for $D_{\rm 3d}=0.1\,\mu{\rm m}^2/{\rm ms}$, $D_{\rm 1d}=10^{-4}\,\mu{\rm m}^2/{\rm ms}$, $r_{\rm int}=3\, {\rm nm}$ and $k_{\rm off}^{\rm ns}=1\,{\rm ms}^{-1}$. The black solid line is the full numerical result based on Eqs. (\ref{int}), (\ref{eq:konk1d}) and (\ref{eq:Wbulkcyl}), while the red dashed line matching at low $k_{\rm on}^{\rm ns}$ is the result in Eqs. (\ref{eq:reactlimit1}) \& (\ref{eq:konk1d}) and the blue dash-dotted line is the result in Eq. (\ref{antenna}).}
\label{fig:koncyl}
\end{figure}

We are now in a position where we can evaluate the integral in Eq. (\ref{int}) numerically, since we can obtain $\lambda_{\rm bulk}$ from $W_{\mathrm{bulk}}^{\rm cyl}$ simply by setting $u=0$, i.e., ${\bar q}=|q|$. The numerical result for $k_{\rm on}$ is shown as a function of $k^{\rm ns}_{\rm on}$ in Fig. \ref{fig:koncyl}, together with two asymptotic approximations. The asymptotic expression for small $k^{\rm ns}_{\rm on}$ is based on Eqs. (\ref{eq:reactlimit1}) and (\ref{eq:konk1d}) and is limited by the non-specific binding to DNA. This limit holds when $k^{\rm ns}_{\rm on}\ll D_{\rm 3d}$. In the opposite limit, $k^{\rm ns}_{\rm on}\gg D_{\rm 3d}$, the detailed form of $W_{\mathrm{bulk}}^{\rm cyl}(q,u=0)$ at small values of $|q|$ becomes important. To obtain an expression for $k_{\rm on}$ in this limit, we use that the asymptotic behaviour of the modified Bessel functions at small arguments are: $K_1(x)\approx 1/x$ and $K_0(x)\approx -\ln(x)$. Thus at small $|q|$, $W_{\mathrm{bulk}}^{\rm cyl}(q,u=0)$ will deviate from unity only by a small term that decays slowly as $-1/\ln(|q|r_{\rm int})$. If we ignore this logarithmic factor and examine the behaviour of the denominator of the integrand in Eq. (\ref{int}), we see that the slow logarithmic behaviour becomes dominant at $|q|$ below $1/l_{\rm sl}^{\rm eff}$ where
\begin{equation}
\label{eff_slide}
l_{\rm sl}^{\rm eff}=\sqrt{\frac{k_{\rm on}^{\rm ns}}{(2\pi D_{\rm 3d})}} l_{\rm sl}.
\end{equation}
If we simply replace the last term in the denominator by its asymptotic value at $q=1/l_{\rm sl}^{\rm eff}$, i.e., we set $1-\lambda_{\rm bulk}(q)= 2\pi D_{\rm 3d}/[k_{\rm on}^{\rm ns}\ln(l_{\rm sl}^{\rm eff}/r_{\rm int})]$ then we can perform the integral and obtain
\begin{equation}
k_{\rm on}\sim \frac{4 \pi D_{\rm 3d} l_{\rm sl}^{\rm eff}}{[\ln(l_{\rm sl}^{\rm eff}/r_{\rm int})]^{1/2}}
\label{antenna}
\end{equation}
This is the approximate asymptotic expression plotted for large $k_{\rm on}^{\rm ns}$ in Fig. \ref{fig:koncyl}. The crossover to this asymptotic behaviour occurs around $k_{\rm on}^{\rm ns}\approx 2\pi D_{\rm 3d}$, but the final convergence only sets in  very slowly as $k_{\rm on}^{\rm ns}$ increases due to the involved logarithmic behaviours.

In the limit when $k_{\rm on}^{\rm ns} \gg D_{\rm 3d}$ the proteins that fall of the DNA will have a tendency to rebind immediately, thus performing what is in effect just a tiny hop on the spot. The tiny probability that the protein avoids such a hop must be proportional to $D_{\rm 3d}/k_{\rm on}^{\rm ns}$ by dimensional arguments. Therefore the number of hops and thus the time spent loosely bound on the DNA scales with $k_{\rm on}^{\rm ns}/D_{\rm 3d}$. The effective distance covered by one-dimensional diffusion along the DNA while sliding and hopping on the spot will therefore increase by a factor proportional to $\sqrt{k_{\rm on}^{\rm ns}/D_{\rm 3d}}$ since the displacement in Brownian motion increase by the square root of time. With this reasoning an interpretation becomes evident of $l_{\rm sl}^{\rm eff}$ as the effective sliding length resulting from combining sliding events connected by quick rebindings. By comparing Eq. (\ref{antenna}) with the Smoluchowski result (\ref{eq:smolres}) this reasoning also provides an interpretation of $l_{\rm sl}^{\rm eff}$ as an increased target size (sometimes referred to as an antenna \cite{hu06}) with $1/[\ln(l^{\rm eff}_{\rm sl}/r_{\rm int})]^{1/2}$ entering as a geometrical factor due to the cylindrical shape of the target.

Finally, we note that the results of Eq. (\ref{int}) and (\ref{eq:Wbulkcyl}) match the result obtained by Berg and Ehrenberg in 1982 \cite{berg82} for infinitely long straight DNA.

\section{Coiled DNA}
\label{sec:coiled}

\begin{figure}[t]
\begin{center}
\includegraphics[width=6.0cm]{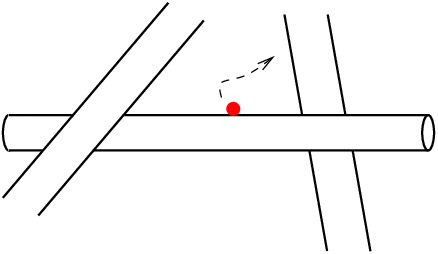}
\quad\quad\quad\quad
\includegraphics[width=3.8cm]{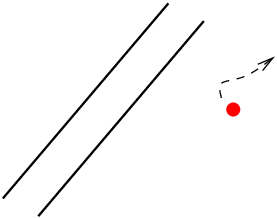}
\end{center}
\caption{Left: The DNA is considered made up of locally straight rods with the protein diffusing among them. Right: The diffusion problem is solved for each rod individually and then combined assuming independent interactions.}
\label{fig:coil}
\end{figure}

We now turn to the calculation of $k_{\rm on}$ when the DNA is coiled. To
reach an analytical expression, we assume that the DNA has a sufficiently
high persistence length, such that locally around the target the DNA can
effectively be thought of as a number of cylindrical straight rods passing
close to each other (see left part of Fig. \ref{fig:coil}). The idea behind
this is that we can solve the diffusion problem for a single straight rod (see
right part of Fig. \ref{fig:coil}), and then we can combine the solutions by
assuming independent interactions. In this independent interactions approach,
the individual solutions are combined by letting the actual binding event
be the binding event that happens first among the multiple rods. This means
that we can calculate the ``survival'' probability of the protein not
having bound anywhere as simply the product of all the probabilities for
each rod that it has not yet bound to that individual rod.

We call $P_{\rm surv}^{\rm foreign}(t)$ the survival probability that the protein has not yet bound to any foreign straight rod ignoring the existence of the (non-foreign) straight rod that it unbound from. With the assumption of independent interactions we can write
\begin{equation}
P_{\rm surv}^{\rm foreign}(t)=\left[1-J_{\rm single}(t)\right]^N,
\label{eq:combforeign}
\end{equation}
where $N$ is the, for the moment finite, number of foreign straight rods and $J_{\rm single}(t)$ is the probability that it has reacted with a particular foreign rod ignoring the existence of all the others. Assuming we have found $J_{\rm single}(t)$ and thereby $P_{\rm surv}^{\rm foreign}(t)$ we will then calculate $W_{\rm bulk}$ as
\begin{equation}
W_{\rm bulk}(q,t)=W_{\rm bulk}^{\rm cyl}(q,t)P_{\rm surv}^{\rm foreign}(t)\quad (q\ne 0),
\label{eq:coilmodel}
\end{equation}
where $W_{\rm bulk}^{\rm cyl}$ is the solution found in the last section when the DNA was a single straight rod. Here we have again used the assumption of independent interactions to calculate the probability that it returns to the straight rod it unbound from, i.e., we multiply the solution without foreign rods with the probability that it has not yet bound to any foreign rod. Also, we have assumed that binding to a foreign rod leads the protein so far away from where it unbound that these foreign rod-bindings do not contribute significantly to the integral of Eq. (\ref{int}), and we have therefore discarded them in Eq. (\ref{eq:coilmodel}). This is the same kind of approximation that was used previously when arriving at the reaction-limited result in Eq. (\ref{eq:reactlimit1}). We have indicated the approximation by the condition $q\ne 0$ in Eq. (\ref{eq:coilmodel}), since these excursions, no matter how long they tend to be, will always contribute at $q=0$ to $W_{\rm bulk}(q,t)$. 

The single foreign rod binding probability $J_{\rm single}$ will be found by solving the usual three-dimensional diffusion equation, Eq. (\ref{eq:bulkdiff}), but now with the initial condition that the protein can start anywhere in the bulk relative to the rod. This means that the initial condition is a uniform concentration, $P(x,r,t=0)=1/V$, with $V$ being a constant. For finite $V$ this implies that $P(x,r,t=0)$ is not normalized. However, $V$ will later be taken to infinity in a way where it will represent the volume of the bulk water, thus restoring proper normalization of probability when this limit is taken. After a Laplace transformation of time $t$ we get
\begin{equation}
u P(r,u)-\frac{1}{V}=D_{\rm 3d}\frac{1}{r}\frac{\partial}{\partial r}r\frac{\partial}{\partial r}{P}(r,u),
\label{eq:capturediff}
\end{equation}
where the $x$ dependence has been left out since the problem is uniform along the $x$ axis. The general solution to this equation is
\begin{equation}
    P(r,u)=\frac{1}{uV}+A I_0(\sqrt{u/D_{\rm 3d}} r)+B K_0(\sqrt{u/D_{\rm 3d}} r),
\end{equation}
with the integration constants $A$ and $B$ that may depend on $u$. As one of the boundary conditions we take that the probability density $P$ should approach the constant $1/V$ far from the DNA. In Laplace space this boundary condition can be written as $\lim_{r\to \infty} P(r,u)=1/(uV)$. Thus we have that $A=0$. For the boundary condition on the DNA surface we employ Eq. (\ref{eq:bulkbound}) without the release from the surface, i.e.,
\begin{equation}
-\left. D_{\rm 3d}\frac{\partial {P}}{\partial r}\right|_{r=r_{\rm int}}=-\left.\frac{k^{\rm ns}_{\rm on}}{2 \pi r_{\rm int}}{ P}\right|_{r=r_{\rm int}}.
\end{equation}
We then obtain
\begin{equation}
   P(r,u)=\frac{1}{uV}-\frac{k_{\rm on}^{\rm ns} K_0(\sqrt{u/D_{\rm 3d}} r)/(u V)}{k^{\rm ns}_{\rm on} K_0(\sqrt{u/D_{\rm 3d}}r_{\rm int}) + 2 \pi r_{\rm int} \sqrt{D_{\rm 3d} u} K_1({\bar q} r_{\rm int})}.
\end{equation}
We can now find the probability of having reacted with a single rod, $J_{\rm single}(t)=\int_0^t L k_{\rm on}^{\rm ns} P(r_{\rm int},t')dt'$, which in Laplace space becomes
\begin{equation}
   J_{\rm single}(u)=\frac{L k_{\rm on}^{\rm ns}}{u^2 V}\left(1+\frac{k_{\rm on}^{\rm ns} K_0(\sqrt{u/D_{\rm 3d}}r_{\rm int})}{2\pi\sqrt{u D_{\rm 3d}}r_{\rm int} K_1(\sqrt{u/D_{\rm 3d}}r_{\rm int})}\right)^{-1}.
\end{equation}

This puts us in a position to combine the results for each rod via Eq. (\ref{eq:combforeign}), and take the limit of $N$, $L$ and $V$ growing to infinity with $L/V=l_{\rm DNA}/N$ for constant $l_{\rm DNA}$. This results in
\begin{equation}
P_{\rm surv}^{\rm foreign}(t)=\exp\left[-J_{\rm cap}(t)\right],
\end{equation}
where $J_{\rm cap}(t)$ is the inverse Laplace transform of
\begin{equation}
J_{\rm cap}(u)= \frac{k_{\rm on}^{\rm ns} l_{\rm DNA}}{u^2}\left(1+\frac{k_{\rm on}^{\rm ns} K_0(\sqrt{u/D_{\rm 3d}}r_{\rm int})}{2\pi\sqrt{u D_{\rm 3d}}r_{\rm int} K_1(\sqrt{u/D_{\rm 3d}}r_{\rm int})}\right)^{-1}.
\label{eq:Jcapu}
\end{equation}

\begin{figure}[t]
\begin{center}
\includegraphics[width=8cm]{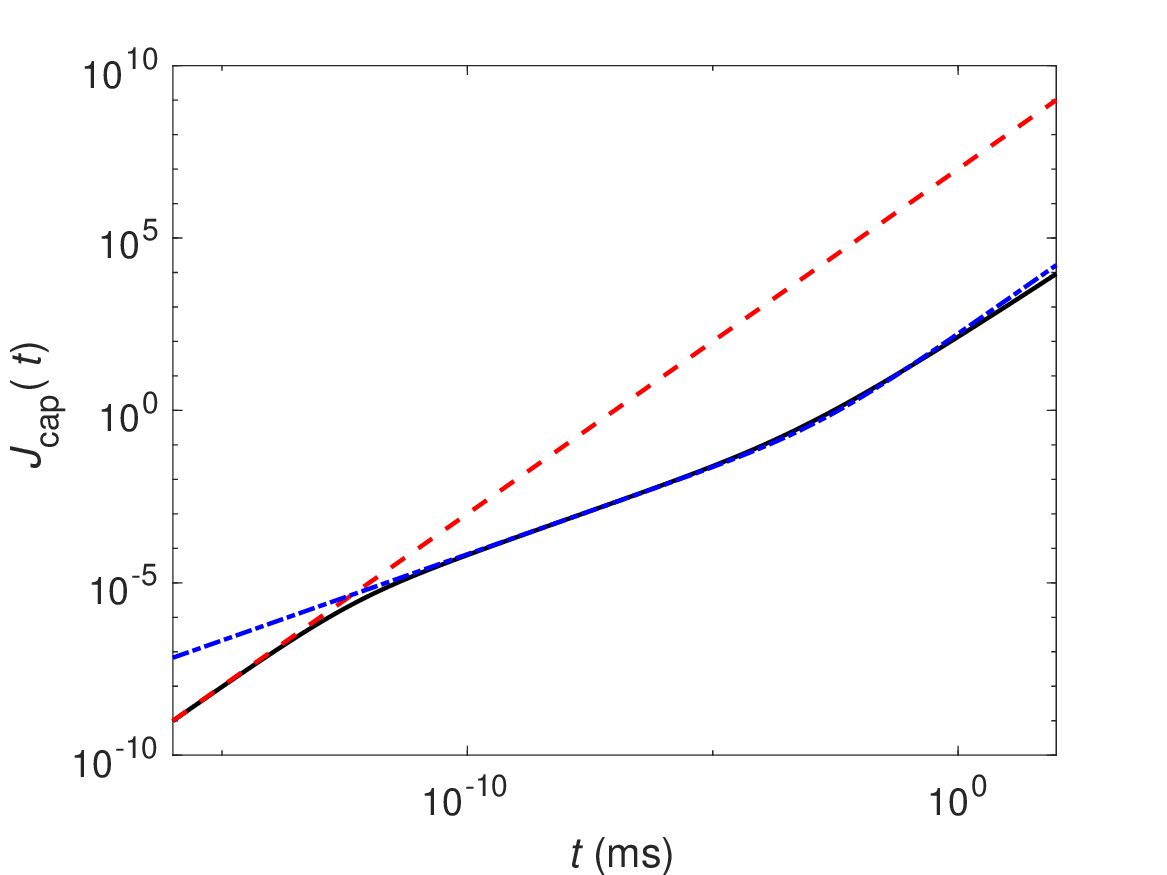}
\end{center}
\caption{Plot of $J_{\rm cap}(t)$ as a function of time $t$ for $D_{\rm 3d}=0.1\,\mu{\rm m}^2/{\rm ms}$, $r_{\rm int}=3\, {\rm nm}$ and $k_{\rm on}^{\rm ns}=10^4 \,\mu{\rm m}^2/{\rm ms}$. The black solid line is the result of numerically Laplace inverting Eq. (\ref{eq:Jcapu}), while the red dashed line matching at low $t$ is the result for the first regime in Eq. (\ref{eq:Jcaptregimes}) and the blue dash-dotted line is the approximation $J_{\rm app}(t)$ given in Eq. (\ref{eq:Papprox}). }
\label{fig:Jcap}
\end{figure}

We can evaluate the inverse Laplace transform of $J_{\rm cap}(u)$ numerically and thereby obtain $P_{\rm surv}^{\rm foreign}(t)$. However, to obtain $W_{\rm bulk}(q,u)$ we would have to perform a Laplace transform on top of that. To reduce the number of transforms, i.e., integrals, that need to be performed we will use a convenient approximation for $J_{\rm cap}(t)$. To argue for this approximation we first note that for $k^{\rm ns}_{\rm on}\gg D_{\rm 3d}$ we can divide the $u$-dependence of $J_{\rm cap}(u)$ into the three regimes
\begin{equation}
J_{\rm cap}(u)\approx\left\{\begin{array}{ll} k^{\rm ns}_{\rm on}l_{\rm DNA}{u^{-2}}&,\; u^{-1}\ll \frac{D_{\rm 3d}r_{\rm int}^2}{(k^{\rm ns}_{\rm on})^2}\\
2\pi l_{\rm DNA}r_{\rm int}\sqrt{D_{\rm 3d}}u^{-3/2}&,\;\frac{D_{\rm 3d}r_{\rm int}^2}{(k^{\rm ns}_{\rm on})^2}\ll u^{-1}\ll \frac{r_{\rm int}^2}{D_{\rm 3d}}\;.\\
\frac{4\pi l_{\rm DNA}D_{\rm 3d}}{u^2\ln (D_{\rm 3d}/(ur_{\rm
int}^2))} &,\;u^{-1}\gg \frac{r_{\rm int}^2}{D_{\rm 3d}}
\end{array}\right.
\end{equation}
When the regimes are well separated in this way, with decaying $u$ dependence, they can each be Laplace inverted independently to find
\begin{equation}
J_{\rm cap}(t)\approx\left\{\begin{array}{ll} k^{\rm ns}_{\rm on}l_{\rm DNA} t&,\; t\ll \frac{D_{\rm 3d}r_{\rm int}^2}{(k^{\rm ns}_{\rm on})^2}\\
4l_{\rm DNA}r_{\rm int}\sqrt{\pi D_{\rm 3d}t}&,\;\frac{D_{\rm 3d}r_{\rm int}^2}{(k^{\rm ns}_{\rm on})^2}\ll t\ll \frac{r_{\rm int}^2}{D_{\rm 3d}}\;,\\
\frac{4\pi l_{\rm DNA}D_{\rm 3d} t}{\ln (D_{\rm 3d}t/r_{\rm int}^2)} &,\;t\gg \frac{r_{\rm int}^2}{D_{\rm 3d}}
\end{array}\right.
\label{eq:Jcaptregimes}
\end{equation}
where Tauberian theorems \cite{hughes95} were applied in the last regime. When $k^{\rm ns}_{\rm on}\gg D_{\rm 3d}$ the first regime will be very brief and the probability of being captured by a foreign rod will only accumulate to $l_{\rm DNA}r_{\rm int}^2 D_{\rm 3d}/k^{\rm ns}_{\rm on}$. This will be a very small number, and therefore we will ignore this regime. In the last regime we will approximate the slowly varying logarithm with a constant. Thus we will assume $J_{\rm cap}(t)\approx k_{\rm cap} t$ in this regime where the precise value of $k_{\rm cap}$ will be fixed later, but it will be somewhere around $l_{\rm DNA}D_{\rm 3d}$. Combining the last two regimes we see that a function that gives the correct approximate behaviour in each is
\begin{equation}
P_{\rm app}(t)=\exp[-J_{\rm app}(t)]\;\;\mathrm{with}\;\; J_{\rm app}(t)=4 l_{\rm DNA} r_{\rm int} \sqrt{\pi D_{\rm 3d} t}+k_{\rm cap}t,
\label{eq:Papprox}
\end{equation}
and this is the function we will replace $P_{\mathrm{surv}}^{\mathrm{foreign}}$ with. To fix $k_{\rm cap}$ we choose the condition that we want the same probability that the protein returns to the non-foreign rod for both of $P_{\mathrm{surv}}^{\mathrm{foreign}}$ and $P_{\mathrm{app}}$. This gives the following equation for $k_{\rm cap}$
\begin{eqnarray}
&&\int_0^\infty d t\;W_{\rm bulk}^{\rm cyl}(q=0,t)\left[P_{\rm surv}^{\rm foreign}(t)
-P_{\rm app}(t)\right]=0,
\end{eqnarray}
which can be handled numerically. This equation involves two Laplace inversions and an integral over these for each numerical guess at a solution, but we have a good guess by using $l_{\rm DNA}D_{\rm 3d}$ as a starting value for $k_{\rm cap}$. See Fig. \ref{fig:Jcap} for a comparison of $J_{\rm cap}(t)$ and $J_{\rm app}(t)$. The numerical Laplace inversions were handled using the Talbot and Euler algorithms presented in \cite{abate06}.

The $\sqrt{t}$ part of $P_{\rm app}$ can be rewritten as a Laplace transform in the following way
\begin{eqnarray}
P_{\rm app}(t)&=&\int_0^\infty d s\; e^{-(s+k_{\rm cap}) t}\frac{a}{2\sqrt{
\pi s^3}}e^{-a^2/(4 s)},
\end{eqnarray}
where $a=4\sqrt{\pi}r_{\rm int}l_{\rm DNA}\sqrt{D_{\rm 3d}}$. This is very convenient when it is inserted
in Eq. (\ref{eq:coilmodel}) instead of $P_{\mathrm{surv}}^{\mathrm{foreign}}$ and Laplace transformed to obtain $W_{\rm bulk}(q,u)$. Then one obtains
\begin{equation}
\label{res}
W_{\rm bulk}(q,u)=\int_0^\infty d s\;\frac{a\, e^{-a^2/(4 s)}}{2\sqrt{\pi s^3}}
W_{\rm bulk}^{\rm cyl}(q,u+s+k_{\rm cap}).
\end{equation}
With this expression the number of involved integrals have been reduced to a single one, and the expression is now ready to be used inside the integral of Eq. (\ref{int}). Along the way in the derivation of the expression we assumed $D_{\rm 3d}\ll k_{\rm on}^{\rm ns}$. But note that in the opposite limit
$D_{\rm 3d}\gg k_{\rm on}^{\rm ns}$ the expression is consistent with the vanishing of $\lambda_{\rm bulk}(q)$ for $q\ne 0$ and thus it extrapolates to the result in Eq. (\ref{eq:reactlimit1}) in that limit. 

To get some analytic insight into the effects of the coiling or presence of foreign DNA we can apply approximations similar to the ones taken in the $D_{\rm 3d}\ll k_{\rm on}^{\rm ns}$ regime of the straight rod. Thus we approximate $W_{\rm bulk}^{\rm cyl}$ by setting $q=1/l_{\rm sl}^{\rm eff}$, and furthermore we approximate $k_{\rm cap}=D_{\rm 3d}l_{\rm DNA}$. Using this we can write
\begin{align}
\label{approx1}
1-\lambda_{\rm bulk}(q)&\approx\int_0^\infty d s\;\frac{a\, e^{-a^2/(4 s)}}{2\sqrt{\pi s^3}}[1-
W_{\rm bulk}^{\rm cyl}(q=1/l_{\rm sl}^{\rm eff},u=s+k_{\rm cap})]\nonumber\\
&=C_1+C_2.
\end{align}
In the second line we split the integral over $s$ in two parts, where $C_1$ represents the integral from $s=0$ to $s=D_{\rm 3d}/r_{\rm int}^2$ and $C_2$ the remaining integral to infinity.

To obtain an approximation for $C_1$ we assume that the dominant contribution of the integral occurs for $s\ll k_{\rm cap}+D_{\rm 3d}/(l_{\rm sl}^{\rm eff})^{2}$ and thus we simply ignore the $s$ dependence in $W_{\rm bulk}^{\rm cyl}$. This gives
\begin{align}
C_1 &\approx\int_0^{D_{\rm 3d}/r_{\rm int}^2} d s\;\frac{a\, e^{-a^2/(4 s)}}{2\sqrt{\pi s^3}} [1-W_{\rm bulk}^{\rm cyl}(q=1/l_{\rm sl}^{\rm eff},u=k_{\rm cap})]\nonumber\\
&\approx 1-W_{\rm bulk}^{\rm cyl}(q=1/l_{\rm sl}^{\rm eff},u=k_{\rm cap})\nonumber\\
&\approx -\frac{2\pi D_{\rm 3d}}{k_{\rm on}^{\rm ns} \ln\left(\sqrt{(l_{\rm sl}^{\rm eff})^{-2}+l_{\rm DNA}}r_{\rm int}\right)}.
\end{align}
For the approximation in the second line we used $a^2\ll D_{\rm 3d}/r_{\rm int}^2$ to extend the integral back to infinity again, and normalisation of the factor in front of the square brackets to perform the integration. In the third line we used the asymptotic expansion of the modified Bessel functions at small arguments, as we did when deriving Eq. (\ref{antenna}).

To obtain an approximation for $C_2$ we do the opposite as for $C_1$ and ignore instead $1/l_{\rm sl}^{\rm eff}$ and $k_{\rm cap}$ for the remainder of the integral. This gives 
\begin{align}
C_2 &\approx\int_{D_{\rm 3d}/r_{\rm int}^2}^\infty d s\;\frac{a\, e^{-a^2/(4 s)}}{2\sqrt{\pi s^3}} [1-W_{\rm bulk}^{\rm cyl}(q=0,u=s)]\nonumber\\
&\approx \int_{D_{\rm 3d}/r_{\rm int}^2}^{\infty} d s\, \frac{a}{2\sqrt{\pi s^3}}\frac{1}{1+k_{\rm on}^{\rm ns}/(2\pi r_{\rm int}\sqrt{D_{\rm 3d} s})}\nonumber\\
&=4 r_{\rm int}^2 l_{\rm DNA}\frac{2\pi D_{\rm 3d}}{k_{\rm on}^{\rm ns}}\ln\left(1+\frac{k_{\rm on}^{\rm ns}}{2\pi D_{\rm 3d}}\right).
\end{align}
For the approximation in the second line we used $a^2\ll D_{\rm 3d}/r_{\rm int}^2$ to discard the exponential and we used that the asymptotic expansion of the two modified Bessel functions are identical for ${\bar q}r_{\rm int}\gg 1$ to discard these functions too. 

Finally, setting $1-\lambda_{\rm bulk}(q)=C_1+C_2$ we can perform the $q$ integral in Eq. (\ref{int}) and reach the approximate expression
\begin{equation}
k_{\rm on}\sim 4 \pi D_{\rm 3d} l_{\rm sl}^{\rm eff}\sqrt{\frac{k_{\rm on}^{\rm ns}}{2\pi D_{\rm 3d}}(C_1+C_2)}.
\label{eq:approxres}
\end{equation}

\begin{figure}[t]
%\sidecaption
\begin{center}
\includegraphics[width=8cm]{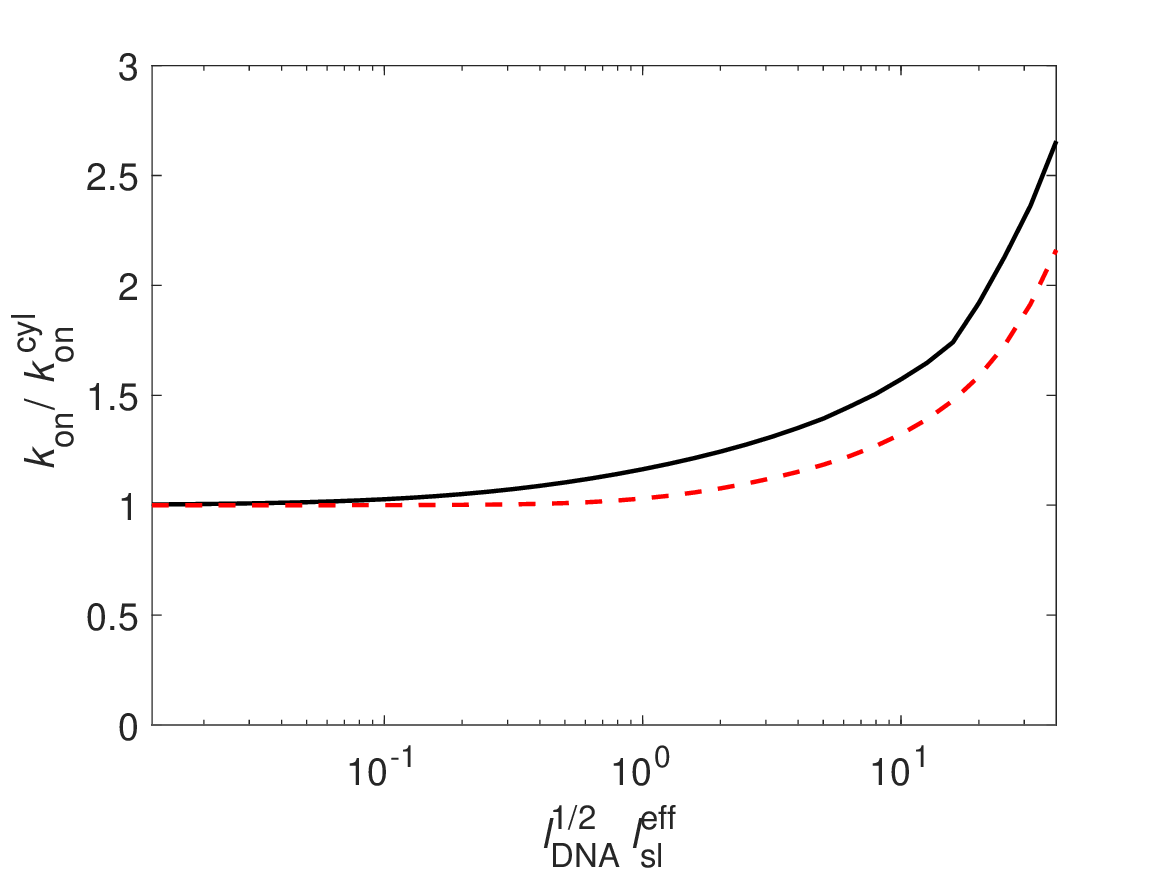}
\end{center}
\caption{Plot of $k_{\rm on}/k_{\rm on}^{\rm cyl}$ as a function of $\sqrt{l_{\rm DNA}}l_{\rm sl}^{\rm eff}$ for $D_{\rm 3d}=0.1\,\mu{\rm m}^2/{\rm ms}$, $D_{\rm 1d}=10^{-4}\,\mu{\rm m}^2/{\rm ms}$, $r_{\rm int}=3\, {\rm nm}$, $k_{\rm on}^{\rm ns}=10^4\,\mu{\rm m}^2/{\rm ms}$ and $k_{\rm off}^{\rm ns}=1\,{\rm ms}^{-1}$ (giving $l_{\rm sl}^{\rm eff}=1.26\,\mu{\rm m}$). $k_{\rm on}^{\rm cyl}$ represents the value for straight DNA, i.e., for $l_{\rm DNA}=0$. The black solid line is the numerical result obtained using Eq. (\ref{res}) in Eq. (\ref{int}), while the red dashed line is the result in Eq. (\ref{eq:approxres}) divided by the result in Eq. (\ref{antenna}). The value of $k_{\rm on}^{\rm ns}$ has been chosen to be far inside the regime $k_{\rm on}^{\rm ns}\gg D_{\rm 3d}$.}
\label{fig:ratios}
\end{figure}

In Fig. \ref{fig:ratios} the ratio of $k_{\rm on}$ for the coiled and straight DNA is plotted as a function of $l_{\rm DNA}^{1/2}$ for both using Eq. (\ref{res}) in Eq. (\ref{int}) and for the approximate result of Eq. (\ref{eq:approxres}) divided by the asymptotic straight approximation in Eq. (\ref{antenna}). The approximate result shows clear deviations, but it qualitatively captures the steep increase in $k_{\rm on}$ when $l_{\rm DNA}$ becomes denser than $1/(l_{\rm sl}^{\rm eff})^{2}$.

To give a feeling for the quantity $l_{\rm DNA}$ we provide an estimate of it in the special case of the Worm Like Chain model. According to \cite{ringrose99} the probability that a point on the DNA a contour distance $s$ from the target happens to be in a tiny volume $d V$ around the target can be approximated by the expression
\begin{equation}
j_{\rm M}(s)d V = \left(\frac{3}{4\pi|s|\ell_{\rm p}}\right)^{3/2}\exp\left(
-\frac{\beta\ell_{\rm p}^2}{|s|^2}\right) d V ,\label{eq:WLC}
\end{equation}
where $\ell_{\rm p}$ is the persistence length of the DNA and $\beta\approx 8$.
Integrating this over the full DNA contour we obtain the average amount of DNA in $d V$, i.e., we obtain the DNA density
\begin{equation}
l_{\rm DNA} = \int_{-L_1}^{L_2} j_{\rm M}(s) d s,
\end{equation}
where $L_1$ and $L_2$ are the lengths of the DNA before and after the target. If we take the limit of infinity long DNA, i.e., $L_1$ and $L_2$ being infinite, we find
\begin{equation}
l_{\rm DNA}=\left(\frac{3}{4\pi}\right)^{3/2}\beta^{-1/4}\Gamma(1/4)\, \ell_{\rm p}^{-2}\approx 0.25\, \ell_{\rm p}^{-2}
\end{equation}
using $\beta\approx 8$. For a persistence length around $\ell_{\rm p}\approx 50\,{\rm nm}$ \cite{manning06} this provides us with the length scale of $l_{\rm DNA}^{-1/2}\approx 100\,{\rm nm}$ for the distances to nearby DNA segments.

\section{Intersegmental transfers}
\label{sec:intra}

Some proteins, such as the lac repressor \cite{hippel89}, are able to bind to more than one DNA strand simultaneously. As a consequence these proteins can move directly from one segment of DNA to another, without having to unbind from the DNA and diffuse temporarily in the bulk water. This mechanism is called an intersegmental transfer \cite{hippel89}, and it can be incorporated in the diffusion equation along the DNA, Eq. (\ref{eq:themodel}), by introducing the extra term
\begin{eqnarray}
&&\hspace*{-0.4cm}\frac{\partial n(x,t)}{\partial t}=\left(D_{\rm 1d}\frac{\partial^2}{ \partial x^2}-k^{\rm ns}_{\rm off}\right)n(x,t)-j(t)\delta(x)+G(x,t)\nonumber\\
&&+k^{\rm ns}_{\rm off}\int_{-\infty}^\infty d x'\int_0^t d t'\,W_{\rm bulk}(x-x',t-t')n(x',t')\nonumber\\
&&-k_{\rm intra}\left(n(x,t)-\int_{-\infty}^\infty d x'\,\lambda_{\rm intra}(x-x')n(x',t)\right).
\label{eq:intra}
\end{eqnarray}
Here $k_{\rm intra}$ is the rate at which intersegmental transfers are
occurring and $\lambda_{\rm intra}(x)$ is the probability density of distances the
protein transfers measured along the DNA. We here again assume independence
between transfers, i.e., that the dynamics of the conformation of
the coil is fast compared with the rate of the transfers $k_{\rm intra}$,
to simplify the modelling considerably. By applying the same mathematical
steps as in Section \ref{sec:facilitated} we find that the rate $k_{\rm 1d}$
can now be expressed as
\begin{equation}
\label{eq:intra2}
k_{\rm 1d}^{-1}=\int_{-\infty}^\infty \frac{d q}{2\pi}
\frac{1}{D_{\rm 1d}q^2+k_{\rm off}^{\rm ns}(1-\lambda_{\rm bulk}(q))+k_{\rm intra}(1-\lambda_{\rm intra}(q))}.
\end{equation}
We conclude from this, that the intersegmental transfers will increase
$k_{\rm 1d}$ and result in faster target search, since the extra term
contributes positively in the denominator of the integrand. Thus increasing
$k_{\rm intra}$ will always result in faster target search. This is opposed
to increasing $k_{\rm off}^{\rm ns}$, since here there will be the opposing
effects of increasing $k_{\rm 1d}$ but lowering the density of loosely bound
proteins $n_{\rm ns}^{\rm eq}$ on the DNA.

There are two limits in which we can simplify the form of the extra term in
Eq. (\ref{eq:intra2}). One occurs when the transfers are typically very long,
so that they do not lead the protein to a place it has already visited. In
this case we can make the approximation $\lambda_{\rm intra}(q)\approx 0$
that we have previously used for $\lambda_{\rm bulk}$ also. This limit
has been studied in \cite{sheinman09}, and it was found there that the
intersegmental transfers can have a significant impact with respect
to lowering search times. Another relatively simple limit, studied in
\cite{lomholt05search}, can emerge if the polymer that the protein diffuses
along is a freely fluctuating coil with low persistence length (i.e., for
DNA with a persistence length of around 150 base pairs \cite{manning06}, this
would require a coarse-graining in a sufficiently long chain). In this
case the distribution $\lambda_{\rm intra}(x)$ can have a power law tail
at large $x$, i.e., $\lambda_{\rm intra}(x)\propto |x|^{-1-\alpha}$ ($x$
large), where for instance $\alpha=0.5$ for Gaussian chains. Motions with
power law tails for jumping distances with $0<\alpha<2$ are called L{\'e}vy
flights. If the power law tail sets in at distances short enough such that
the motion is dominated by one-dimensional diffusion along the chain at these
distances, then it will only be the power law tail of the intersegmental
transfer term that contributes to the integral in Eq. (\ref{eq:intra2}),
and we can approximate $k_{\rm intra}(1-\lambda_{\rm intra}(q))\approx
D_{\rm L}|q|^\alpha$ with $D_{\rm L}$ being a diffusion constant for the
L{\'e}vy flight. This $q$-scaling means that the corresponding convolution
term in Eq. (\ref{eq:intra}) can be written as a fractional derivative. See
\cite{lomholt05search} for more on this including a study of optimal search
strategies, i.e., values of $k_{\rm off}^{\rm ns}$ that minimises search
time in this limit.

\section{Conclusion}
\label{sec:conclusion}

The mathematical theory presented in this chapter provides a way to
make predictions about the effect of coiling of DNA for target search
by proteins. For proteins with only one DNA binding site the effect of
coiling is quantified through the concentration $l_{\rm DNA}$ of DNA around
the local segment of DNA with the target. However, the theory relies on many assumptions that limit
the applications. These include: the DNA molecule is long and has a
large persistence length, no correlations with previous bulk excursions
(possibly through dynamics of the DNA), and dilute protein concentration. Thus
there are many relevant alternatives of modelling the effect of DNA coiling on
target search that has not been covered in this chapter. Just a few examples
are: scaling regimes when the persistence length cannot be assumed to be
long \cite{hu06}, effect of large scale heterogeneous network structure
\cite{nyberg21,hedstrom23}, or effect of the fractal nature of chromatin
\cite{benichou11}.

Sequence heterogeneity was mentioned in Section \ref{sec:facilitated} as a
potential barrier for fast search. However, combined with looping, sequence
heterogeneity in the form of auxiliary sites within looping distance of the
target, can also help in accelerating search \cite{bauer15}. Without looping
such sites will act as traps, and this introduces additional long-time scales
in the search process. Modelling such traps will mean that the assumption
of non-specific equilibration and exponentially distributed search time
employed in this chapter will have to be examined carefully \cite{benichou09}.

We finish by noting that more recent studies have shown that at low chemical
concentrations chemical rate constants (i.e., inverse mean search times) may
become insufficient. Instead, even in quite generic geometries, repeated
first-passage times may be significantly different from each other
\cite{monasterio11,mattos12}, with characteristic times spanning several orders of
magnitude \cite{godec16,grebenkov18}. Given that transcription factors in living
cells may reach such low numbers (ten or few tens in the entire cell), it
will be of interest to explore the full span of relevant target localisation
times, in supercomputing studies \cite{ma20} or experimentally. In such scenarios
of low copy numbers the bursty protein production of a DNA-binding proteins
encoded in one gene leads to strong concentration fluctuations at the target
gene, in situations when the gene-gene distance is short \cite{pulkkinen13}.

Facilitated diffusion was demonstrated to occur both in vitro and in living
cells. Single-DNA experiments unveiled contributions of DNA coiling to the
localisation dynamics of DNA-binding proteins to their specific target site
on the DNA. In the future it will be of interest to follow individual
proteins from their production \cite{yu06} across the cell \cite{rienzo14}, to
their eventual binding to the DNA.

\bibliographystyle{apsrev4-1}
%\nocite{apsrev41Control}


\end{document}